\documentclass[12pt]{article}
\usepackage{epsf}
\usepackage{amsmath,amssymb}

\usepackage[dvips]{graphicx}

\usepackage{comment}

\begin{document}

\newcommand{\lsim}{\stackrel{<}{_\sim}}
\newcommand{\gsim}{\stackrel{>}{_\sim}}
\newcommand{\mathhyphen}{\mathchar"712D}

\newcommand{\rem}[1]{{$\spadesuit$\bf #1$\spadesuit$}}

\renewcommand{\thefootnote}{\fnsymbol{footnote}}
\setcounter{footnote}{0}

\begin{titlepage}

\begin{center}

\hfill KEK-TH-1964\\
\hfill UT-17-09\\
\hfill March, 2017

\vskip .5in

{\Large\bf
On the Gauge Invariance\\
of the Decay Rate of False Vacuum\\
}

\vskip .5in

{\large
  Motoi Endo$^{\rm (a,b,c)}$, Takeo Moroi$^{\rm (d,c)}$, 
  Mihoko M. Nojiri$^{\rm (a,b,c)}$, Yutaro Shoji$^{\rm (e)}$
}

\vskip 0.25in

$^{\rm (a)}${\em 
KEK Theory Center, IPNS, KEK, Tsukuba, Ibaraki 305-0801, Japan}

\vskip 0.1in
$^{\rm (b)}${\em 
The Graduate University of Advanced Studies (Sokendai),\\
Tsukuba, Ibaraki 305-0801, Japan}

\vskip 0.1in
$^{\rm (c)}${\em 
Kavli IPMU (WPI), University of Tokyo, Kashiwa, Chiba 277-8583, Japan}

\vskip 0.1in
$^{\rm (d)}${\em 
Department of Physics, University of Tokyo, Tokyo 113-0033, Japan}

\vskip 0.1in
$^{\rm (e)}${\em 
Institute for Cosmic Ray Research, The University of Tokyo, 
Kashiwa 277-8582, Japan}

\end{center}

\vskip .3in

\begin{abstract}

  We study the gauge invariance of the decay rate of the false vacuum
  for the model in which the scalar field responsible for the false
  vacuum decay has gauge quantum number.  In order to calculate the
  decay rate, one should integrate out the field fluctuations around
  the classical path connecting the false and true vacua (i.e.,
  so-called bounce).  Concentrating on the case where the gauge
  symmetry is broken in the false vacuum, we show a systematic way to
  perform such an integration and present a manifestly gauge-invariant
  formula of the decay rate of the false vacuum.

\end{abstract}

\end{titlepage}

\setcounter{page}{1}
\renewcommand{\thefootnote}{\#\arabic{footnote}}
\setcounter{footnote}{0}

There have been continuous interest in the theoretically correct
calculation of the decay rate of the false vacuum.  One of the recent
motivations has been provided by the discovery of the Higgs boson at
the LHC \cite{Aad:2015zhl} and the precision measurement of the top
quark mass at the LHC and Tevatron \cite{ATLAS:2014wva}; in the
standard model, we are facing the possibility to live in a metastable
electroweak vacuum with lifetime much longer than the age of the
universe \cite{Isidori:2001bm, Degrassi:2012ry, Alekhin:2012py,
  Espinosa:2015qea, Plascencia:2015pga, Lalak:2016zlv,
  Espinosa:2016nld}.  Furthermore, the false and true vacua may show
up in various models of physics beyond the standard model.  One
important example is supersymmetric standard model in which the
electroweak symmetry breaking vacuum may become unstable with the
existence of the color or charge breaking vacuum at which colored or
charged sfermion fields acquire vacuum expectation values; the
condition that the electroweak vacuum has sufficiently large lifetime
constrains the parameters in supersymmetric models
\cite{Gunion:1987qv, Casas:1995pd, Kusenko:1996jn, Hisano:2010re,
  Camargo-Molina:2013sta, Chowdhury:2013dka, Blinov:2013fta,
  Camargo-Molina:2014pwa, Endo:2015oia, Endo:2015ixx}.  Thus, detailed
understanding of the decay of the false vacuum is important in
particle physics and cosmology.

In \cite{Coleman:1977py, Callan:1977pt, Coleman:aspectsof}, the
calculation of the decay rate of the false vacuum was formulated with
the so-called bounce configuration which is a solution of the
4-dimensional (4D) Euclidean equation of motion connecting false
vacuum and true vacuum (more rigorously, the other side of the
potential wall).  The decay rate of the false vacuum per unit volume
is given in the following form:
\begin{align}
  \gamma = {\cal A} e^{-{\cal B}},
  \label{decayrate}
\end{align}
where ${\cal B}$ is the bounce action, while the prefactor ${\cal A}$
is obtained by integrating out field fluctuations around the bounce
configuration as well as those around the false vacuum.

In gauge theories, if a scalar field with gauge quantum number
acquires non-vanishing amplitude at the true or false vacuum, the
gauge, Higgs and the ghost sectors contribute to ${\cal A}$.  The
decay rate should be calculated with the gauge-fixed Lagrangian which
contains the gauge parameter $\xi$.  In the present study, we
concentrate on the gauge dependence (i.e., the $\xi$-dependence) of
the decay rate of the false vacuum.  Formally, the $\xi$-dependence of
${\cal A}$ should cancel out exactly.  This is due to the fact that
the decay rate is derived from the effective action of the bounce
configuration, and also that the effective action for any solution of
the equation of motion is assured to be gauge invariant
\cite{Nielsen:1975fs, Fukuda:1975di}.  In the actual calculation,
however, the gauge independence is not manifest because the
$\xi$-dependence should cancel out among the contributions of gauge
field, Nambu-Goldstone (NG) boson, and Faddeev-Popov (FP)
ghosts.\footnote
{The gauge invariance of the effective potential of the model we
  consider was discussed in \cite{Alexander:2008hd}; however, the
  scalar configuration was assumed to be space-time independent, and
  hence the result is not applicable to the present case.  The gauge
  independence of the sphaleron transition rate was studied in
  \cite{Baacke:1999sc} using functional determinant method which is
  also adopted in our analysis.}
In particular, the gauge boson and the NG mode, whose fluctuation
operator is $\xi$-dependent, mix with each other around the bounce
configuration.  This makes the study of the decay rate complicated.
Furthermore, it is difficult to check the gauge independence even
numerically because a stable numerical implementation proposed so far
requires $\xi=1$.

In this letter, we show a procedure to integrate out the field
fluctuations, which gives rise to a manifestly gauge invariant
expression of the decay rate overcoming the difficulties mentioned
above.  In the current study, we concentrate on the case where
\begin{enumerate}
\item the gauge symmetry is $U(1)$,\footnote
  {The application of our prescription to the case of non-abelian
    gauge symmetry is straightforward.}
\item there is only one charged scalar field $\Phi$ which affects the
  decay of the false vacuum,
\item the $U(1)$ symmetry is spontaneously broken in the false vacuum.
\end{enumerate}
More general cases, in particular, the case where the $U(1)$ symmetry
is preserved at the false vacuum, is discussed in \cite{Endo:2017tsz}.

First, let us explain the set up of our analysis. The Euclidean
Lagrangian is given by
\begin{align}
  {\cal L} = \frac{1}{4} F_{\mu\nu} F_{\mu\nu} 
  + [ (\partial_\mu + i g A_\mu) \Phi^\dagger ]
  [ (\partial_\mu - i g A_\mu) \Phi ]
  + V + {\cal L}_{\rm G.F.} + {\cal L}_{\rm ghost},
  \label{Ltot}
\end{align}
where $A_\mu$ is the gauge field, $F_{\mu\nu}=\partial_\mu
A_\nu-\partial_\nu A_\mu$, and $V$ is the scalar potential.  In
addition, ${\cal L}_{\rm G.F.}$ and ${\cal L}_{\rm ghost}$ are the
gauge-fixing term and the terms containing FP ghosts (denoted as $c$
and $\bar{c}$), respectively.  We use the following gauge-fixing
function:\footnote
{Previous studies used different type of the gauge-fixing functions:
  $\partial_\mu A_\mu - \sqrt{2} \xi g \bar{\phi} {\rm Re} \Phi$,
  around the bounce (i.e., $\Phi=\bar{\phi}/\sqrt{2}$), and
  $\partial_\mu A_\mu - \sqrt{2} \xi g v {\rm Im}\Phi$, around the
  false vacuum (i.e., $\Phi=v/\sqrt{2}$).  Expanding the fields around
  the solution of the classical equation of motion, we obtain the same
  gauge-fixing functions as the previous studies at least at the
  one-loop level, although our gauge-fixing function can be used both
  around the bounce and around the false vacuum.}
\begin{align}
  {\cal F} = \partial_\mu A_\mu - 2 \xi g ({\rm Re} \Phi) ({\rm Im} \Phi)
  = \partial_\mu A_\mu 
  + \frac{i}{2} \xi g ( \Phi^2 - {\Phi^\dagger}^2 ),
  \label{gaugefixingfn}
\end{align}
with which 
\begin{align}
  {\cal L}_{\rm G.F.} = \frac{1}{2\xi} {\cal F}^2,
\end{align}
and
\begin{align}
  {\cal L}_{\rm ghost} =
  \bar{c} 
  \left[
    -\partial_\mu \partial_\mu 
    + \xi g^2 ( \Phi^2 + {\Phi^\dagger}^2 )
  \right] c.
\end{align}

The scalar potential $V$ has true and false vacua. We assume that the
true and false vacua exist at the tree-level; we do not consider the
case where the second vacuum is radiatively generated.  The field
configuration of the false vacuum is expressed as\footnote
{The field amplitude at the false vacuum (as well as the bounce
  configuration) may be shifted due to loop effects; the shifts are
  $\xi$-dependent in general.  However, at the one-loop level, the
  shifts do not affect the extremum values of the effective action to
  which the decay rate of the false vacuum is related.}
\begin{align}
  (A_\mu,\Phi)_{\rm false\, vacuum}
  =(0,v/\sqrt{2}),
\end{align}
with $v$ being a constant which is non-vanishing in this letter.

The false vacuum decay is dominated by the classical path, so-called
the bounce \cite{Coleman:1977py}.  When $v\neq 0$, the bounce
solution, which is $O(4)$ symmetric \cite{Coleman:1977th,
  Blum:2016ipp}, is given in the following form:
\begin{align}
  (A_\mu,\Phi)_{\rm bounce} = (0, \bar{\phi}(r)/\sqrt{2}),
  \label{BounceSolution}
\end{align}
where $r\equiv\sqrt{x_\mu x_\mu}$ is the radius of the 4D Euclidean
space.  Here, the function $\bar{\phi}$ is a solution of the classical
equation of motion:
\begin{align}
  \left[
    \partial_r^2 \Phi + \frac{3}{r} \partial_r \Phi
    - V_\Phi
  \right]_{\Phi\rightarrow\bar{\phi}/\sqrt{2}}
  = 0,
  \label{ClassicalEq}
\end{align}
where $V_\Phi$ denotes the derivative of the scalar potential with
respect to $\Phi$.  It also satisfies the following boundary
conditions:
\begin{align}
  &\partial_r \bar{\phi} (r=0) = 0,
  \\
  &\bar{\phi} (r=\infty) = v.
\end{align}
We assume that $\bar{\phi}$ is a real function of $r$.  At
$r\rightarrow\infty$, $\bar{\phi}$ settles on the false-vacuum; in
such a limit, $\bar{\phi}$ (approximately) obeys the following
equation:
\begin{align}
  \partial_r^2 \bar{\phi} + \frac{3}{r} \partial_r \bar{\phi} -
  m_h^2 (\bar{\phi}-v) \simeq 0,
\end{align}
where $m_h$ is the mass of the (massive) scalar boson around the
false vacuum.  Then, the asymptotic behavior of $\bar{\phi}$ can be
expressed as
\begin{align}
  \bar{\phi} (r\rightarrow\infty) 
  \simeq v + \kappa \frac{e^{-m_h r}}{r^{3/2}},
  \label{barphi(inf)}
\end{align}
with $\kappa$ being a constant.

For the calculation of the decay rate of the false vacuum, it is
necessary to integrate out the fluctuations around the bounce.  The
gauge and scalar fields are decomposed around the bounce as
\begin{align}
  A_\mu = a_\mu,~~~
  \Phi = \frac{1}{\sqrt{2}}
  \left( \bar{\phi} + h + i \varphi \right),
\end{align}
where the ``Higgs'' mode $h$ and the ``NG'' mode $\varphi$ are real
fields.  We expand the field fluctuations as\footnote
{For notational simplicity, we omit the subscripts $J$, $m_A$, and
  $m_B$ from the radial function $\alpha$'s, and the summations over
  $J$, $m_A$, and $m_B$ are implicit.}
\begin{align}
  a_\mu (x) \ni &\,
  \alpha_S (r) \frac{x_\mu}{r} {\cal Y}_{J,m_A,m_B}
  + \alpha_L (r) \frac{r}{L} \partial_\mu {\cal Y}_{J,m_A,m_B}
  \nonumber \\ &\, 
  + \alpha_{T1} (r) 
  i \epsilon_{\mu\nu\rho\sigma} V^{(1)}_\nu L_{\rho\sigma} {\cal Y}_{J,m_A,m_B}
  + \alpha_{T2} (r) 
  i \epsilon_{\mu\nu\rho\sigma} V^{(2)}_\nu L_{\rho\sigma} {\cal Y}_{J,m_A,m_B},
  \\
  h (x) \ni &\, \alpha_h (r) {\cal Y}_{J,m_A,m_B},
  \\
  \varphi (x) \ni &\, \alpha_\varphi (r) {\cal Y}_{J,m_A,m_B}.
\end{align}
where ${\cal Y}_{J,m_A,m_B}$ denotes the 4D hyperspherical harmonics;
the eigenvalues of $S_A^2$, $S_B^2$, $S_{A,3}$ $S_{B,3}$ (with $S_A$
and $S_B$ being generators of the rotational group of the 4D Euclidean
space, i.e., $SU(2)_A\times SU(2)_B$) are $J(J+1)$, $J(J+1)$, $m_A$,
and $m_B$, respectively.  Notice that $J=0$, $\frac{1}{2}$, $1$,
$\cdots$.  In addition, $V^{(1)}_\nu$ and $V^{(2)}_\nu$ are
(arbitrary) two independent vectors,
$L_{\rho\sigma}\equiv\frac{i}{\sqrt{2}}
(x_\rho\partial_\sigma-x_\sigma\partial_\rho)$, and
\begin{align}
  L \equiv \sqrt{4J (J +1)}.
\end{align}

For $J>0$, the fluctuation operator for
$(\alpha_S,\alpha_L,\alpha_\varphi)$ is obtained as
\begin{align}
  {\cal M}_J^{(S,L,\varphi)} \equiv &
  \left(
    \begin{array}{ccc}
      \displaystyle{ -\Delta_J + \frac{3}{r^2} + g^2 \bar{\phi}^2 }
      & \displaystyle{ -\frac{2L}{r^2} } & 2 g \bar{\phi}' 
      \\[3mm]
      \displaystyle{ -\frac{2L}{r^2} } & 
      \displaystyle{ -\Delta_J - \frac{1}{r^2} + g^2 \bar{\phi}^2 }
      & 0
      \\[3mm]
      2 g \bar{\phi}' & 0 
      & \displaystyle{ 
        -\Delta_J + \frac{(\Delta_0 \bar{\phi})}{\bar{\phi}} + \xi g^2 \bar{\phi}^2 }
    \end{array}
  \right)
  \nonumber \\[2mm] & + 
  \left( 1 - \frac{1}{\xi} \right)
  \left(
    \begin{array}{ccc}
      \displaystyle{ \partial_r^2 + \frac{3}{r} \partial_r - \frac{3}{r^2} }
      & \displaystyle{ 
        -L \left(\frac{1}{r} \partial_r - \frac{1}{r^2} \right) } & 0 
      \\[3mm]
      \displaystyle{ L \left( \frac{1}{r} \partial_r + \frac{3}{r^2} \right) } 
      & \displaystyle{ -\frac{L^2}{r^2} } & 0
      \\[3mm]
      0 & 0 & 0
    \end{array}
  \right),  
  \label{M_J}
\end{align}
where $\bar{\phi}'\equiv\partial_r\bar{\phi}$, and
\begin{align}
  \Delta_J \equiv \partial_r^2 + \frac{3}{r} \partial_r 
  - \frac{L^2}{r^2}.
\end{align}
For $J=0$, $\alpha_L$-mode does not exist, and the fluctuation
operator is in $2\times 2$ form as
\begin{align}
  {\cal M}_{J=0}^{(S,\varphi)} \equiv &
  \left(
    \begin{array}{cc}
      \displaystyle{ 
        \frac{1}{\xi} \left( 
          -\Delta_0 + \frac{3}{r^2} + \xi g^2 \bar{\phi}^2 
        \right) }
      &
      2 g \bar{\phi}'
      \\[3mm]
      2 g \bar{\phi}'
      &
      \displaystyle{ 
        -\Delta_0 + \frac{(\Delta_0 \bar{\phi})}{\bar{\phi}}
        + \xi g^2 \bar{\phi}^2 }
    \end{array}
  \right).
  \label{M_J=0}
\end{align}
In addition, the fluctuation operator for the transverse modes, the
Higgs mode, and the FP ghost mode are given by
\begin{align} 
&\,  {\cal M}_J^{(T)} = -\Delta_J + g^2 \bar{\phi}^2
\label{MT}
\\ & \, 
  {\cal M}_J^{(h)} = 
  \left[ 
    -\Delta_J + V_{\Phi\Phi^\dagger} 
  \right]_{\Phi\rightarrow\bar{\phi}/\sqrt{2}},
  \label{MHiggs}
\\ &\,
  {\cal M}_J^{(\bar{c},c)} = -\Delta_J + \xi g^2 \bar{\phi}^2,
  \label{MFP}
\end{align}
with
$V_{\Phi\Phi^\dagger}\equiv\partial^2V/\partial\Phi\partial\Phi^\dagger$. 

We also need the fluctuation operators around the false vacuum,
denoted as $\widehat{\cal M}_J^{(S,L,\varphi)}$, $\widehat{\cal
  M}_J^{(T)}$, and so on.  (Here and hereafter, the ``hat'' is used
for objects related to the false vacuum.)  They can be obtained from
the fluctuation operators Eqs.\ \eqref{M_J}, \eqref{M_J=0},
\eqref{MT}, \eqref{MHiggs}, and \eqref{MFP} by replacing
$\bar{\phi}\rightarrow v$, and $\bar{\phi}'\rightarrow 0$.

The prefactor ${\cal A}$ in Eq.\ \eqref{decayrate} is related to the
functional determinants of the fluctuation operators introduced above.
It can be expressed as \cite{Callan:1977pt}
\begin{align}
  {\cal A} = \frac{{\cal B}^2}{4\pi^2} 
  {\cal A}'^{(h)}
  {\cal A}^{(S, L, \varphi)}
  {\cal A}^{(T)} 
  {\cal A}^{(\bar{c},c)} 
  {\cal A}^{(\rm extra)},
  \label{PrefactorA}
\end{align}
where ${\cal A}'^{(h)}$, ${\cal A}^{(S,L,\varphi)}$, ${\cal A}^{(T)}$,
and ${\cal A}^{(\bar{c},c)}$ are contributions of the Higgs mode,
$(\alpha_S,\alpha_L,\varphi)$, $(\alpha_{T_1},\alpha_{T2})$, and FP
ghosts, respectively, which are given by
\begin{align}
  {\cal A}'^{(h)} = &
  \left[
    \frac{{\mbox{Det}'} {\cal M}_{1/2}^{(h)}}
    {\mbox{Det} \widehat{\cal M}_{1/2}^{(h)}}
  \right]^{-2}
  \prod_{J\neq 1/2}
  \left[
    \frac{\mbox{Det} {\cal M}_J^{(h)}}
    {\mbox{Det} \widehat{\cal M}_J^{(h)}}
  \right]^{-(2J +1)^2/2},
  \label{A'(Higgs)}
  \\
  {\cal A}^{(S,L,\varphi)} = &\,
  \left[
    \frac{\mbox{Det} {\cal M}_0^{(S,\varphi)}}
    {\mbox{Det} \widehat{\cal M}_0^{(S,\varphi)}}
  \right]^{-1/2}
  \prod_{J=1/2}^{\infty} \left[
    \frac{\mbox{Det} {\cal M}_J^{(S,L,\varphi)}}
    {\mbox{Det} \widehat{\cal M}_J^{(S,L,\varphi)}}
  \right]^{-(2J +1)^2/2},
  \\
  {\cal A}^{(T)} = &\,
  \prod_{J=1/2}^{\infty} \left[
    \frac{\mbox{Det} {\cal M}_J^{(T)}}
    {\mbox{Det} \widehat{\cal M}_J^{(T)}}
  \right]^{-(2J +1)^2},
  \\
  {\cal A}^{(\bar{c},c)} = &\,
  \prod_{J=0}^{\infty} \left[
    \frac{\mbox{Det} {\cal M}_J^{(\bar{c},c)}}
    {\mbox{Det} \widehat{\cal M}_J^{(\bar{c},c)}}
  \right]^{(2J +1)^2}.
  \label{A(ghost)}
\end{align}
Here, ``prime'' in Eq.\ \eqref{A'(Higgs)} indicates that the effect of
the zero modes in association with the translational invariance is
omitted in calculating the functional determinant
\cite{Callan:1977pt}.  The contributions of extra fields other than
those introduced above are expressed by ${\cal A}^{(\rm extra)}$; we
do not consider them in this letter.  We are interested in the gauge
dependence of the decay rate, therefore we focus on the $S$, $L$, and
NG modes as well as FP ghosts whose fluctuation operators are
dependent on $\xi$.

Our main task is to calculate the functional determinants mentioned
above.  For this purpose, we use the method discussed in
\cite{Coleman:aspectsof, Dashen:1974ci, Kirsten:2003py,
  Kirsten:2004qv}.  With $N\times N$ fluctuation operators ${\cal
  M}^{(X)}$ and $\widehat{\cal M}^{(X)}$ being given, we introduce two
sets of $N$ linearly independent functions $\psi^{(X)}_I$ and
$\widehat{\psi}^{(X)}_I$ ($I=1-N$), obeying ${\cal M}^{(X)}\psi_I=0$
and $\widehat{\cal M}^{(X)}\widehat{\psi}_I=0$.  Here, $\psi_I$ and
$\widehat{\psi}_I$ satisfy the same boundary condition at $r=0$.
Then, the ratio of the functional determinants is related to their
asymptotic behaviors at $r\rightarrow\infty$ as
\begin{align}
  \frac{\mbox{Det} {\cal M}^{(X)}}{\mbox{Det} \widehat{\cal M}^{(X)}}
  = \lim_{r\rightarrow\infty}
  \frac{\mbox{det}(\psi_1(r)~\cdots~\psi_N(r))}
  {\mbox{det}(\widehat{\psi}_1(r)~\cdots~\widehat{\psi}_N(r))}.
\end{align}
In the following, we use the above relation to evaluate the functional
determinants of the fluctuation operators given in Eqs.\
\eqref{A'(Higgs)} $-$ \eqref{A(ghost)}.  For our study, $\psi_I$ and
$\widehat{\psi}_I$ are required to be regular at $r=0$ for the
finiteness of the effective action.

The fluctuation operator for the ghost is given in Eq.\ \eqref{MFP}.
For the calculation of its functional determinant, we define
the function $f^{\rm (FP)}_J$ which obeys
\begin{align}
  (\Delta_J - \xi g^2 \bar{\phi}^2) f^{\rm (FP)}_J = 0,
  \label{eq_fFP}
\end{align}
where the boundary condition of $f^{\rm (FP)}_J$ is taken to be
\begin{align}
  f^{\rm (FP)}_J (r\rightarrow 0) \simeq r^{2J}.
\end{align}
We also introduce the function $\widehat{f}^{\rm (FP)}_J$ which obeys
\begin{align}
  (\Delta_J - \xi g^2 v^2) \widehat{f}^{\rm (FP)}_J = 0,
\end{align}
with
\begin{align}
  \widehat{f}^{\rm (FP)}_J (r\rightarrow 0) \simeq r^{2J}.
\end{align}
The explicit form of $\widehat{f}^{\rm (FP)}_J$ is given by
\begin{align}
  \widehat{f}^{\rm (FP)}_J (r) = 2^{2J + 1} \Gamma (2 J + 2)
  (gv)^{-(2J +1)}
  \frac{I_{2J +1} (\sqrt{\xi} g v r)}{r},
\end{align}
where $I_{2J +1}$ is the modified Bessel function.  Then,
\begin{align}
  \frac{\mbox{Det} {\cal M}_J^{(\bar{c},c)}}
  {\mbox{Det} \widehat{\cal M}_J^{(\bar{c},c)}}
  = 
  \frac{f^{\rm (FP)}_J (r\rightarrow\infty)}
  {\widehat{f}^{\rm (FP)}_J (r\rightarrow\infty)}.
\end{align}

For the contributions of the $S$-, $L$-, and $\varphi$-modes with
$J>0$, we need the functions $\Psi$ and $\widehat{\Psi}$, which are
regular at the origin, satisfying
\begin{align}
  &
  {\cal M}^{(S,L,\varphi)}_J \Psi = 0,
  \label{Ml*Psi=0}
  \\ &
  \widehat{\cal M}^{(S,L,\varphi)}_J \widehat{\Psi} = 0.
\end{align}
Hereafter, the boundary conditions for $\Psi$ and $\widehat{\Psi}$ at
the origin are taken to be the same.  With three independent solutions
of the above equations (which we denote $\Psi_I$ and
$\widehat{\Psi}_I$, with $I=1$, $2$, and $3$), the functional
determinants of our interests are given by
\begin{align}
  \frac{\mbox{Det} {\cal M}_J^{(S,L,\varphi)}}
  {\mbox{Det} \widehat{\cal M}_J^{(S,L,\varphi)}} = 
  \frac{ {\cal D}_J^{(S,L,\varphi)} (r\rightarrow\infty)}
  { \widehat{\cal D}_J^{(S,L,\varphi)} (r\rightarrow\infty)},
  \label{detM/detMhat}
\end{align}
where
\begin{align}
  {\cal D}^{(S,L,\varphi)}_J (r) \equiv &\,
  \mbox{det}
  (\Psi_1 (r) ~ \Psi_2 (r) ~ \Psi_3 (r)),
  \label{det}
  \\
  \widehat{\cal D}^{(S,L,\varphi)}_J (r) \equiv &\,
  \mbox{det}
  (\widehat{\Psi}_1 (r) ~ \widehat{\Psi}_2 (r) ~ 
  \widehat{\Psi}_3 (r)).
\end{align}

Hereafter, we use the fact that the solution of Eq.\ \eqref{Ml*Psi=0}
can be decomposed as
\begin{align}
  \Psi 
  \equiv
  \left( \begin{array}{c}
      \Psi^{\rm (top)} \\
      \Psi^{\rm (mid)} \\
      \Psi^{\rm (bot)}
    \end{array} 
  \right) 
  =
  \left( \begin{array}{c}
      \partial_r \chi 
      \\[3mm]
      \displaystyle{\frac{L}{r}} \chi
      \\[3mm] g
      \bar{\phi} \chi
    \end{array} 
  \right) 
  + 
  \left( \begin{array}{c}
      \displaystyle{ \frac{1}{r g^2 \bar{\phi}^2} \eta }
      \\[3mm]
      \displaystyle{ \frac{1}{L r^2 g^2 \bar{\phi}^2} \partial_r (r^2 \eta) }
      \\[3mm] 
      0
    \end{array} 
  \right)
  + 
  \left( \begin{array}{c}
      \displaystyle{ -2 \frac{\bar{\phi}'}{g^2 \bar{\phi}^3} \zeta }
      \\[3mm]
      0 
      \\[3mm] 
      \displaystyle{ \frac{1}{g \bar{\phi}} \zeta }
    \end{array} 
  \right),
  \label{Psi}
\end{align}
where the functions $\chi$, $\eta$, and $\zeta$ obey the following
equations:
\begin{align}
  &
  (\Delta_J - \xi g^2 \bar{\phi}^2) \chi = 
  \frac{2 \bar{\phi}'}{r g^2 \bar{\phi}^3} \eta 
  + \frac{2}{r^3} \partial_r 
  \left( \frac{r^3 \bar{\phi}'}{g^2 \bar{\phi}^3} \zeta \right),
  \label{eq_chi}
  \\[2mm] &
  (\Delta_J - g^2 \bar{\phi}^2) \eta 
  - \frac{2 \bar{\phi}'}{r^2 \bar{\phi}} \partial_r
  \left(
    r^2 \eta
  \right)
  = - \frac{2 L^2 \bar{\phi}'}{r \bar{\phi}} \zeta,
  \label{eq_eta}
  \\[2mm] &
  (\Delta_J - \xi g^2 \bar{\phi}^2) \zeta = 0.
  \label{eq_zeta}
\end{align}
Then, the following identities hold:\footnote
{At the leading order in fluctuations, Eq.\ \eqref{Psitop'} is
  equivalent to $\alpha_{\cal F}+\xi\zeta=0$, where $\alpha_{\cal F}$
  is the radial mode function of the gauge fixing function, i.e.,
  ${\cal F}(x)\ni\alpha_{\cal F}(r){\cal Y}_{J,m_A,m_B}$.}
\begin{align}
  \partial_r \Psi^{\rm (top)} &\, = 
  - \frac{3}{r} \Psi^{\rm (top)}
  + \frac{L}{r} \Psi^{\rm (mid)} + \xi g^2 \bar{\phi}^2 \chi,
  \label{Psitop'}
  \\
  \partial_r \Psi^{\rm (mid)} &\, = 
  \frac{L}{r} \Psi^{\rm (top)}
  - \frac{1}{r} \Psi^{\rm (mid)} + \frac{1}{L} \eta.
  \label{Psimid'}
\end{align}

Hereafter, we give  three independent solutions $\Psi_I$ ($I=1-3$) of
Eq.\ \eqref{Ml*Psi=0}, and show their boundary conditions at $r=0$.
The solutions can be constructed with the following three sets of the
functions $(\chi_I,\eta_I,\zeta_I)$:
\begin{enumerate}
\item For $\Psi_1$, we take $\eta_1=\zeta_1=0$, and 
  \begin{align}
    \chi_1= f^{\rm (FP)}_J,
  \end{align}
  with which Eqs.\ \eqref{eq_chi}, \eqref{eq_eta} and \eqref{eq_zeta}
  are satisfied.  Then,
  \begin{align}
    \Psi_1 (r\rightarrow 0) \simeq 
     \left( \begin{array}{c}
         2J r^{2J-1} \\ L r^{2J-1} \\ g \bar{\phi}_C r^{2J}
       \end{array}
     \right),
     \label{BC:Psi1}
  \end{align}
  where
  \begin{align}
    \bar{\phi}_C \equiv \bar{\phi} (r=0).
  \end{align}
\item For $\Psi_2$, we can take $\zeta_2=0$, and
  \begin{align}
    \chi_2 (r\rightarrow 0) &\, \simeq -
    \frac{1}{2J g^2 \bar{\phi}_C^2} r^{2J},
    \\
    \eta_2 (r\rightarrow 0) &\, \simeq r^{2J}.
  \end{align}
  Then, 
  \begin{align}
    \Psi_2 (r\rightarrow 0) \simeq 
     \left( \begin{array}{c}
         \displaystyle{ -\frac{(J+1)\xi-J}{2L^2} r^{2J+1} }
         \\[3mm]
         \displaystyle{ -\frac{(J+1)\xi-(J+2)}{4 L (J+1)} r^{2J+1} }
         \\[3mm]
         \displaystyle{ -\frac{1}{2J g \bar{\phi}_C} r^{2J} }
       \end{array}
     \right).
     \label{BC:Psi2}
  \end{align}
\item For $\Psi_3$, we take
  \begin{align}
    \zeta_3 = f_J^{\rm (FP)},
  \end{align}
  while $\chi_3 (r\rightarrow 0)$ and $\eta_3 (r\rightarrow 0)$ are
  both $O(r^{2J+2})$.  The contributions to the top and middle
  components of $O(r^{2J+1})$ vanish, and
  \begin{align}
    \Psi_3 (r\rightarrow 0) \simeq 
    \left( \begin{array}{c}
        O(r^{2J+2}) \\[3mm] O(r^{2J+2}) \\[3mm] 
         \displaystyle{ \frac{1}{g \bar{\phi}_C} r^{2J} }
       \end{array}
     \right).
     \label{BC:Psi3}
  \end{align}
\end{enumerate}

The solutions around the false vacuum, denoted as $\widehat{\Psi}_I$,
satisfy the same boundary conditions at $r\rightarrow 0$ as those of
$\Psi_I$, and obey the following differential equation:
\begin{align}
  \widehat{\cal M}^{(S,L,\varphi)}_J \widehat{\Psi}_I = 0.
  \label{hatM*hatPsi=0}
\end{align}
Notice that the evolution equation of the bottom component of
$\widehat{\Psi}_I$ does not contain the top and middle components and
vice versa.

For the following discussion, it is convenient to define the function
$f^{(\eta)}_J$, which obeys
\begin{align}
  (\Delta_J - g^2 \bar{\phi}^2) f^{(\eta)}_J
  - \frac{2 \bar{\phi}'}{r^2 \bar{\phi}} \partial_r
  \left(
    r^2 f^{(\eta)}_J
  \right)
  = 0,
  \label{eq_feta}
\end{align}
and
\begin{align}
  f^{(\eta)}_J (r\rightarrow 0) \simeq r^{2J}.
\end{align}
(For $J=0$, $f^{(\eta)}_0 (r\rightarrow 0) \simeq 1$.)  We emphasize
here that the function $f^{(\eta)}_J$ is independent of $\xi$.  The
homogeneous solutions of Eqs.\ \eqref{eq_chi} and \eqref{eq_zeta}
(that of Eq.\ \eqref{eq_eta}) are given by $f^{\rm (FP)}_J$
($f^{(\eta)}_J$); thus, in particular, $\eta_2=f^{(\eta)}_J$.  We also
define the function $\widehat{f}^{(\eta)}_J$ which obeys
\begin{align}
  (\Delta_J - g^2 v^2) \widehat{f}^{(\eta)}_J = 0,
\end{align}
with
\begin{align}
  \widehat{f}^{(\eta)}_J (r\rightarrow 0) \simeq r^{2J}.
\end{align}

Next, we consider the mode with $J=0$.  The fluctuation operators
${\cal M}_{J=0}^{(S,\varphi)}$ and $\widehat{\cal
  M}_{J=0}^{(S,\varphi)}$ are in $2\times 2$ form.  For the
calculation of their functional determinants, we need the solutions of
the following equations:
\begin{align}
  &
  {\cal M}_{J=0}^{(S,\varphi)} \Psi = 0,
  \label{M0Psi=0}
  \\ &
  \widehat{\cal M}_{J=0}^{(S,\varphi)} \widehat{\Psi} = 0,
\end{align}
with which
\begin{align}
  \frac{\mbox{Det} {\cal M}_0^{(S,\varphi)}}
  {\mbox{Det} \widehat{\cal M}_0^{(S,\varphi)}} = 
  \frac{ {\cal D}_0^{(S,\varphi)} (r\rightarrow\infty)}
  {\widehat{\cal D}_0^{(S,\varphi)} (r\rightarrow\infty)},
\end{align}
where
\begin{align}
  {\cal D}^{(S,\varphi)}_0 (r) \equiv &\,
  \mbox{det}
  (\Psi_1 (r) ~ \Psi_2 (r)),
  \label{det2x2}
  \\
  \widehat{\cal D}^{(S,\varphi)}_0 (r) \equiv &\,
  \mbox{det}
  (\widehat{\Psi}_1 (r) ~ \widehat{\Psi}_2 (r)).
\end{align}

Solutions of \eqref{M0Psi=0} are given in the following form:
\begin{align}
  \Psi \equiv
  \left( \begin{array}{c}
      \Psi^{\rm (top)} \\
      \Psi^{\rm (bot)}
    \end{array} 
  \right) 
  \equiv
  \left( \begin{array}{c}
      \partial_r \chi 
      \\
      g \bar{\phi} \chi
    \end{array} 
  \right) 
  + 
  \left( \begin{array}{c}
      \displaystyle{ -2 \frac{\bar{\phi}'}{g^2 \bar{\phi}^3} \zeta }
      \\[3mm] 
      \displaystyle{ \frac{1}{g \bar{\phi}} \zeta }
    \end{array} 
  \right),
  \label{Psi_0}
\end{align}
where the functions $\chi$ and $\zeta$ obey Eq.\ \eqref{eq_chi} with
$\eta=0$ and Eq.\ \eqref{eq_zeta}, respectively.  Two independent
solutions of Eq.\ \eqref{M0Psi=0} can be chosen as follows:
\begin{enumerate}
\item For $\Psi_1$, we take $\zeta_1=0$, and 
  \begin{align}
    \chi_1 (r\rightarrow 0) \simeq 1.
  \end{align}
  Then,
  \begin{align}
    \Psi_1 (r\rightarrow 0) \simeq 
    \left( \begin{array}{c}
        \displaystyle{ \frac{1}{4} \xi g^2 \bar{\phi}_C^2 r}
        \\[3mm]
        g \bar{\phi}_C
      \end{array}
    \right).
  \end{align}
\item For $\Psi_2$, we take
  \begin{align}
    \zeta_2 (r\rightarrow 0) \simeq 1, 
  \end{align}
  while $\chi_2 (r\rightarrow 0)$ is $O(r^2)$, with which
  \begin{align}
    \Psi_2 (r\rightarrow 0) \simeq 
    \left( \begin{array}{c}
        O(r^2)
        \\[3mm]
        \displaystyle{ \frac{1}{g\bar{\phi}_C} }
      \end{array}
    \right).
  \end{align}
\end{enumerate}
Notice that $\chi_1=\zeta_2=f^{\rm (FP)}_0$.

With the solutions introduced above, we now discuss the decay rate of
the false vacuum.  For this purpose, we study the asymptotic behaviors
of the solutions at $r\rightarrow\infty$.  First, we consider the
modes with $J\neq 0$.  Each of Eq.\ \eqref{eq_fFP} and \eqref{eq_feta}
has only one growing solution at $r\rightarrow\infty$.  The other
solutions are exponentially suppressed at $r\rightarrow\infty$; those
dumping modes are irrelevant for the following discussion.  At
$r\rightarrow\infty$, $f^{\rm (FP)}_J$ and $f^{(\eta)}_J$ behave as
\begin{align}
  f^{\rm (FP)}_J (r\rightarrow\infty) &\, \simeq c_{\rm FP}
  \frac{e^{\sqrt{\xi}gv r}}{r^{3/2}}
  \left[ 1 + O(r^{-1}) \right],
  \\[2mm] 
  f^{(\eta)}_J (r\rightarrow\infty) &\, \simeq c_\eta
  \frac{e^{gv r}}{r^{3/2}}
  \left[ 1 + O(r^{-1}) \right],
\end{align}
where $c_{\rm FP}$ and $c_\eta$ are constants.

The behaviors of $\chi_I$, $\eta_I$, and $\zeta_I$ can be understood
by using $f^{\rm (FP)}_J$ and $f^{(\eta)}_J$, using the fact that
$\bar{\phi}'$ is exponentially suppressed at $r\rightarrow\infty$ (see
Eq.\ \eqref{barphi(inf)}).  Obviously,
\begin{align}
  \chi_1 (r) = f^{\rm (FP)}_J (r).
\end{align}
Because $\chi_2$ is given by the sum of a homogeneous solution and a
particular solution (which we denote $\delta\chi^{(\eta)}$), 
$\Psi_2$ can be expressed by
\begin{align}
  \chi_2 (r) &\, = 
  a_1 f^{\rm (FP)}_J (r) + \delta\chi^{(\eta)} (r),
  \label{chi2(vneq0)}
  \\
  \eta_2 (r) &\, = 
  f^{(\eta)}_J (r),
\end{align}
where $a_1$ is a constant.  The function $\delta\chi^{(\eta)}$
satisfies the following equation:
\begin{align}
  (\Delta_J - \xi g^2 \bar{\phi}^2) \delta\chi^{(\eta)} = 
  \frac{2 \bar{\phi}'}{r g^2 \bar{\phi}^3} f^{(\eta)}_J.
\end{align}
At $r\rightarrow\infty$, we can see that $\delta\chi^{(\eta)}$ behaves
as
\begin{align}
  \delta\chi^{(\eta)} (r\rightarrow\infty ) \simeq
  - \frac{2 m_h \kappa}{g^2 v^3 [(gv-m_h)^2-\xi g^2v^2] r^{5/2}}
  e^{-m_h r} f^{(\eta)}_J
  + \cdots.
\end{align}
Furthermore, the functions for $\Psi_3$ behave as
\begin{align}
  \chi_3 (r) &\, =
  b_1 f^{\rm (FP)}_J (r)  + b_2 \delta\chi^{(\eta)} (r) + 
  \delta\chi^{(\zeta)} (r),
  \label{chi_3}
  \\
  \eta_3 (r) &\, =
  b_2 f^{(\eta)}_J (r) + \delta\eta^{(\zeta)} (r),
  \label{eta_3}
  \\
  \zeta_3 (r) &\, =
  f^{\rm (FP)}_J (r),
\end{align}
with $b_1$ and $b_2$ constants.  The functions $\delta\chi^{(\zeta)}$
and $\delta\eta^{(\zeta)}$ obey the following equations:
\begin{align}
  &
  (\Delta_J - \xi g^2 \bar{\phi}^2) \delta\chi^{(\zeta)} = 
  \frac{2 \bar{\phi}'}{r g^2 \bar{\phi}^3} \delta\eta^{(\zeta)}
  + \frac{2}{r^3} \partial_r 
  \left( \frac{r^3 \bar{\phi}'}{g^2 \bar{\phi}^3} f^{\rm (FP)}_J \right),
  \\[2mm] &
  (\Delta_J - g^2 \bar{\phi}^2) \delta\eta^{(\zeta)}
  - \frac{2 \bar{\phi}'}{r^2 \bar{\phi}} \partial_r
  \left(
    r^2 \delta\eta^{(\zeta)}
  \right)
  = - \frac{2 L^2 \bar{\phi}'}{r \bar{\phi}} f^{\rm (FP)}_J,
\end{align}
and their asymptotic behaviors are\footnote
{In Eq.\ \eqref{f_chi(zeta)}, we do not explicitly show the effect of
  $\delta\eta^{(\zeta)}$ on $\delta\chi^{(\zeta)}$, because it
  is subdominant.}
\begin{align}
  \delta\chi^{(\zeta)} (r\rightarrow\infty) &\, \simeq 
  \frac{2 (\sqrt{\xi}gv-m_h) \kappa}
  {g^2 v^3 (2\sqrt{\xi}gv-m_h)r^{3/2}} 
  e^{-m_h r} f^{\rm (FP)}_J + \cdots,
  \label{f_chi(zeta)}
  \\[2mm]
  \delta\eta^{(\zeta)} (r\rightarrow\infty) &\, \simeq 
  \frac{2 L^2 m_h \kappa}{v [(\sqrt{\xi}gv-m_h)^2 - g^2 v^2]r^{5/2}}
  e^{-m_h r} f^{\rm (FP)}_J + \cdots.
\end{align}

Using the asymptotic behaviors given above, the determinant defined in
Eq.\ \eqref{det} has the following structure:
\begin{align}
  {\cal D}^{(S,L,\varphi)}_J (r\rightarrow\infty) \simeq
  \mbox{det}
  \left( \begin{array}{ccc}
      O(f^{\rm (FP)}_J) &
      O(r^{-1} f^{(\eta)}_J) &
      O(e^{-m_h r} f^{\rm (FP)}_J)
      \\
      O(r^{-1} f^{\rm (FP)}_J) &
      O(f^{(\eta)}_J) &
      O(e^{-m_h r} f^{\rm (FP)}_J)
      \\
      O(f^{\rm (FP)}_J) &
      O(r^{-5/2} e^{-m_h r} f^{(\eta)}_J) &
      O(f^{\rm (FP)}_J) 
    \end{array}
  \right).
\end{align}
The determinant is dominated by the product of the diagonal elements,
and is given by
\begin{align}
  {\cal D}^{(S,L,\varphi)}_J (r\rightarrow\infty) \simeq
  \frac{1}{L g^3 v^3}
  (\partial_r\chi_1) (\partial_r\eta_2) \zeta_3
  \simeq
  \frac{\sqrt{\xi}}{Lgv}
    {f^{\rm (FP)}_J}^2 f^{(\eta)}_J.
\end{align}
Because $\chi_1$ and $\zeta_3$ obey the same equation as that of the
FP mode while $\eta_2$ is $\xi$-independent, ${\cal
  D}^{(S,L,\varphi)}_J(r\rightarrow\infty)$ has a $\xi$-dependence
which can be cancelled out by the contribution from the FP ghosts.

In order to evaluate $\widehat{\Psi}_I$, we can use the fact that the
upper two components of $\widehat{\Psi}_I$ and the bottom component
are decoupled in the evolution equation given in Eq.\
\eqref{hatM*hatPsi=0}.  We can find
\begin{align}
  \widehat{\Psi}_1 (r\rightarrow \infty) \simeq &\,
  \left( \begin{array}{c}
      \displaystyle{ \partial_r \widehat{f}^{\rm (FP)}_J }
      \\[3mm]
      \displaystyle{ \frac{L}{r} \widehat{f}^{\rm (FP)}_J }
      \\[3mm]
      \displaystyle{ g \bar{\phi}_C \widehat{f}^{\rm (FP)}_J }
    \end{array}
  \right),
  \\[3mm]
  \widehat{\Psi}_2 (r\rightarrow \infty) \simeq &\,
  \left( \begin{array}{c}
      \displaystyle{ - \frac{1}{2 J g^2 \bar{\phi}_C^2} 
        \partial_r \widehat{f}^{\rm (FP)}_J
        + \frac{1}{r g^2 v^2} \widehat{f}^{(\eta)}_J }
      \\[3mm]
      \displaystyle{ - \frac{L}{2 J g^2 \bar{\phi}_C^2 r}
        \widehat{f}^{\rm (FP)}_J
        + \frac{1}{L r^2 g^2 v^2} \partial_r (r^2 \widehat{f}^{(\eta)}_J)
      }
      \\[3mm]
      \displaystyle{ - \frac{1}{2 J g \bar{\phi}_C} \widehat{f}^{\rm (FP)}_J }
    \end{array}
  \right),
  \\[3mm]
  \widehat{\Psi}_3 (r\rightarrow \infty) \simeq &\,
  \left( \begin{array}{c}
      0
      \\[3mm]
      0
      \\[3mm]
      \displaystyle{ \frac{1}{g \bar{\phi}_C} \widehat{f}^{\rm (FP)}_J }
    \end{array}
  \right).
\end{align}
The functional determinant around the false vacuum is given by
\begin{align}
  \widehat{\cal D}^{(S,L,\varphi)}_J (r\rightarrow\infty) \simeq
  \frac{\sqrt{\xi}}{Lg \bar{\phi}_C}
  { \widehat{f}{}_J^{\rm (FP)} }^{2}
  \widehat{f}^{(\eta)}_J.
\end{align}

The discussion for
$J=0$ is similar to that for $J\neq 0$. 
The asymptotic behaviors of $\Psi_I$ are given by
\begin{align}
  \Psi_1 (r\rightarrow \infty) \simeq &\,
  \left( \begin{array}{c}
      \displaystyle{ \partial_r f^{\rm (FP)}_0 }
      \\[3mm]
      \displaystyle{ g v f^{\rm (FP)}_0 }
    \end{array}
  \right),
  \\[3mm]
  \Psi_2 (r\rightarrow \infty) \simeq &\,
  \left( \begin{array}{c}
      O(e^{-m_h r} f^{\rm (FP)}_0)
      \\[3mm]
      \displaystyle{ \frac{1}{gv} f^{\rm (FP)}_0 }
    \end{array}
  \right).
\end{align}
For the false vacuum solutions, $\widehat{\Psi}_I$ ($I=1$ and $2$), we
can use the fact that $\widehat{\Psi}^{\rm (top)}$ and
$\widehat{\Psi}^{\rm (bot)}$ evolve independently.  Requiring that
$\widehat{\Psi}_I$ satisfies the same boundary condition as ${\Psi}_I$
at $r\rightarrow 0$, we obtain
\begin{align}
  \widehat{\Psi}_1 (r\rightarrow \infty) \simeq &\,
  \left( \begin{array}{c}
      \displaystyle{ \frac{\bar{\phi}_C^2}{v^2} 
        \partial_r \widehat{f}^{\rm (FP)}_0 }
      \\[3mm]
      \displaystyle{ g \bar{\phi}_C \widehat{f}^{\rm (FP)}_0 }
    \end{array}
  \right),
  \\[3mm]
  \widehat{\Psi}_2 (r\rightarrow \infty) \simeq &\,
  \left( \begin{array}{c}
      0
      \\[3mm]
      \displaystyle{ \frac{1}{g \bar{\phi}_C} \widehat{f}^{\rm (FP)}_0 }
    \end{array}
  \right).
\end{align}
The functional determinant around the bounce and that around the false
vacuum are given by
\begin{align}
  {\cal D}^{(S, \varphi)}_0 (r\rightarrow \infty) \simeq
  \sqrt{\xi} {f_0^{\rm (FP)}}^2,
\end{align}
and
\begin{align}
  \widehat{\cal D}^{(S, \varphi)}_0 (r\rightarrow \infty) \simeq
  \frac{\sqrt{\xi} \bar{\phi}_C}{v}
  {\widehat{f}{}_0^{\rm (FP)}}^2,
\end{align}
respectively.

Combining the effects of the gauge field, the NB boson, and the FP
ghosts, we obtain
\begin{align}
  {\cal A}^{(S,L,\varphi)}
  {\cal A}^{(\bar{c},c)}
  = 
  \left(
    \frac{v}{\bar{\phi}_C}
  \right)^{-1/2}
  \prod_{J\geq 1/2}
  \left[
    \frac{\bar{\phi}_C f^{(\eta)}_J (r\rightarrow\infty)}
    {v \widehat{f}^{(\eta)}_J (r\rightarrow\infty)}
  \right]^{-(2J +1)^2/2}.
  \label{A^LSphi}
\end{align}
We emphasize that the above result is manifestly gauge invariant.  
For completeness, we also summarize the contributions of the
transverse and Higgs modes.  The contribution of transverse mode is 
given by 
\begin{align}
  {\cal A}^{(T)}  =
  \prod_{J=1/2}^{\infty} \left[
    \frac{f^{(T)}_J (r\rightarrow\infty)}
    {\widehat{f}^{(T)}_J (r\rightarrow\infty)}
  \right]^{-(2J +1)^2}.
  \label{A^T}
\end{align}
Here, the functions $f^{(T)}_J$ and $\widehat{f}^{(T)}_J$ satisfy
\begin{align}
  &(\Delta_J - g^2 \bar{\phi}^2) f^{(T)}_J = 0,
  \\
  &(\Delta_J - g^2 v^2) \widehat{f}^{(T)}_J = 0,
\end{align}
with 
\begin{align}
  f^{(T)}_J (r\rightarrow 0) \simeq 
  \widehat{f}^{(T)}_J (r\rightarrow 0) \simeq 
  r^{2J}.
\end{align}
For the Higgs mode contribution, we first define
the functions $f^{(h)}_J$ and $\widehat{f}^{(h)}_J$, satisfying
\begin{align}
  &
  (\Delta_J - V_{\Phi\Phi^\dagger}) f^{(h)}_J = 0,
  \\
  &
  (\Delta_J - m_h^2) \widehat{f}^{(h)}_J = 0,
\end{align}
with 
\begin{align}
  f^{(h)}_J (r\rightarrow 0) \simeq 
  \widehat{f}^{(h)}_J (r\rightarrow 0) \simeq 
  r^{2J}.
\end{align}
As we have mentioned, we need to omit the zero eigenvalues in
association with the translational invariance.  Such zero eigenvalues
show up in the $J=\frac{1}{2}$ mode \cite{Callan:1977pt,
  Endo:2015ixx}, and can be eliminated by using the function
$\check{f}^{(h)}_{1/2}$ which obeys
\begin{align}
  (\Delta_{1/2} - V_{\Phi\Phi^\dagger}) \check{f}^{(h)}_{1/2} = 
  f^{(h)}_{1/2}.
\end{align}
With $\check{f}^{(h)}_{1/2}$, the Higgs mode contribution is given
by
\begin{align}
  {\cal A}'^{(h)} =
  \left[
    \frac{\check{f}^{(h)}_{1/2} (r\rightarrow\infty)}
    {\widehat{f}^{(h)}_{1/2} (r\rightarrow\infty)}
  \right]^{-2}
  \prod_{J\neq 1/2} \left[
    \frac{f^{(h)}_J (r\rightarrow\infty)}
    {\widehat{f}^{(h)}_J (r\rightarrow\infty)}
  \right]^{-(2J +1)^2/2}.
  \label{A^h}
\end{align}

By substituting Eqs.\ \eqref{A^LSphi}, \eqref{A^T}, and \eqref{A^h}
into Eq.\ \eqref{PrefactorA}, we obtain the prefactor ${\cal A}$ for
the calculation of the decay rate of the false vacuum.  With the
present prescription, it is related to the asymptotic behaviors of the
solutions of second-order differential equations which are
$\xi$-independent.  Notice that the formula of the decay rate we
obtained is manifestly gauge invariant.

In summary, in this letter, we have studied the false vacuum decay in
theory with $U(1)$ gauge symmetry, paying particular attention to the
gauge invariance of the decay rate.  Concentrating on the case where
the gauge symmetry is broken in the false vacuum, we derived a
manifestly gauge-invariant expression of the decay rate.  Although we
have studied the case with $U(1)$ gauge symmetry, application of our
result to models with non-abelian gauge groups is straightforward.

We emphasize that our result not only guarantees the gauge invariance
of the decay rate but also simplifies the numerical calculation.  In
order to evaluate the prefactor ${\cal A}$ numerically, one should
calculate the functions $\Psi$, which are three- or two-component
objects, by solving Eq.\ \eqref{Ml*Psi=0} or \eqref{M0Psi=0}.  With a
general value of $\xi$, each mode grows differently at
$r\rightarrow\infty$, and the numerical calculation of the functional
determinant is difficult.  This problem can be avoided if we take the
so-called 't~Hooft-Feynman gauge with $\xi=1$ because, in such a
gauge, the fluctuation operators of the gauge and NG fields become
simple.  Even taking the 't~Hooft-Feynman gauge, however, one has to
solve the coupled equations, which is numerically demanding.  In
addition, if one adopts a particular choice of gauge, like the
't~Hooft-Feynman gauge, the gauge independence of the result is not
explicit.  On the contrary, with our results, only the asymptotic
behaviors of functions which are manifestly $\xi$-independent are
necessary to obtain the decay rate of the false vacuum.  Our simple
formula, which is manifestly gauge independent, would greatly reduce
the numerical costs compared to the previous procedures.

Finally, several comments are in order.
\begin{itemize}
\item When the gauge symmetry is preserved at the false vacuum, i.e.,
  $v=0$, we can find a class of solution of the classical equation of
  motion.  With the function $\bar{\phi}$ obeying Eq.\
  \eqref{ClassicalEq}, the following field configuration satisfies the
  condition for the bounce:
  \begin{align*}
    A_\mu = \frac{1}{g} \partial_\mu \Theta (r), ~~~
    \Phi = \frac{1}{\sqrt{2}} \bar{\phi} (r) e^{i\Theta (r)}.
  \end{align*}
  Here, the function $\Theta$ obeys
  \begin{align*}
    \partial_r^2 \Theta + \frac{3}{r} \partial_r \Theta
    - \frac{1}{2} \xi g^2 \bar{\phi}^2 \sin 2 \Theta = 0,
  \end{align*}
  and $\Theta' (0)=0$ (with the ``prime'' being the derivative with
  respect to $r$).  With such a boundary condition, the function
  $\Theta$ is determined by its value at $r=0$.  In calculating the
  decay rate, we need to take account of the effect of all the
  possible bounce configuration parametrized by $\Theta (0)$.
  Importantly, the fluctuation operator depends on $\Theta (0)$ with
  the present choice of gauge-fixing function, which makes the
  calculation of the decay rate complicated. As hinted in
  \cite{Kusenko:1996bv}, the calculation is simplified with a
  different gauge-fixing function which is independent of the scalar
  field.
\item Related to the previous comment, when the gauge symmetry is
  preserved at the false vacuum, there shows up a zero mode related to
  internal symmetry.  The path integral over such a zero mode should
  be reinterpreted as the integration over the possible bounce
  configuration parametrized by $\Theta (0)$.
\item What we are calculating is the one-loop effective action of the
  bounce, therefore renormalization is necessary.  In other words, in
  the calculation of the prefactor ${\cal A}$, contribution from each
  angular momentum $J$ is finite, but the infinite sum of those
  contributions diverges.  The divergences should be subtracted by
  including the counter terms \cite{Callan:1977pt}.
\end{itemize}
We discuss these issues in a separate publication \cite{Endo:2017tsz}.

\vspace{3mm}
\noindent
{\it Acknowledgements}: This work was supported by the Grant-in-Aid
for Scientific Research on Scientific Research B (No.16H03991 [ME and
MMN] and No.26287039 [MMN]), Scientific Research C (No.26400239 [TM]),
Young Scientists B (No.16K17681 [ME]) and Innovative Areas
(No.16H06490 [TM] and 16H06492 [MMN]), and by World Premier
International Research Center Initiative (WPI Initiative), MEXT,
Japan.

\end{document}